\documentclass[journal=ancac3,manuscript=article]{achemso}
\usepackage{xcolor}

\newcommand{\blue}[1]{\iffalse\color{blue}#1~\color{black}\fi}

\usepackage{bm}% bold math
\usepackage{graphicx}
\usepackage{siunitx}

\DeclareSIUnit\angstrom{\protect \text {Å}}

\title{ Graphene bilayer as a template for manufacturing novel encapsulated 2D materials}

\author{Arkady V. Krasheninnikov}
\affiliation{Institute of Ion Beam Physics and Materials Research,
Helmholtz-Zentrum Dresden-Rossendorf
01328 Dresden, Germany}
\alsoaffiliation{The Institute of Scientific and Industrial Research (ISIR-SANKEN), Osaka University, Osaka 567- 0047, Japan}
\email{a.krasheninnikov@hzdr.de}

\author{Yung-Chang Lin}
\affiliation{The Institute of Scientific and Industrial Research (ISIR-SANKEN), Osaka University, Osaka 567- 0047, Japan}
\alsoaffiliation{Nanomaterials Research Institute, National Institute of Advanced Industrial Science and Technology (AIST), Tsukuba 305-8565, Japan}
\email{yc-lin@aist.go.jp}

\author{Kazu Suenaga}
\affiliation{The Institute of Scientific and Industrial Research (ISIR-SANKEN), Osaka University, Osaka 567- 0047, Japan}
\alsoaffiliation{Nanomaterials Research Institute, National Institute of Advanced Industrial Science and Technology (AIST), Tsukuba 305-8565, Japan}
\email{suenaga-kazu@ sanken.osaka-u.ac.jp}

\date{\today}

\begin{document}

\begin{abstract}
Bilayer graphene (BLG) has recently been used as a tool to stabilize the
encapsulated single sheets of various layered materials and tune their
properties. It was also discovered that the protecting action of
graphene sheets makes it possible to synthesize completely new
two-dimensional materials (2DMs) inside BLG by intercalating various
atoms and molecules. In comparison to the bulk graphite, BLG allows for
easier intercalation and much larger increase in the inter-layer
separation of the sheets. Moreover, it enables studying the atomic
structure of the intercalated 2DM using high-resolution transmission
electron microscopy. In this review, we summarize the recent progress in
this area, with a special focus on new materials created inside BLG. We
compare the experimental findings with the theoretical predictions, pay
special attention to the discrepancies and outline the challenges in the
field. Finally, we discuss unique opportunities oﬀered by the
intercalation into 2DMs beyond graphene and their heterostructures.

\end{abstract}

\begin{figure}
    \centering
    \includegraphics[width=0.5\linewidth]{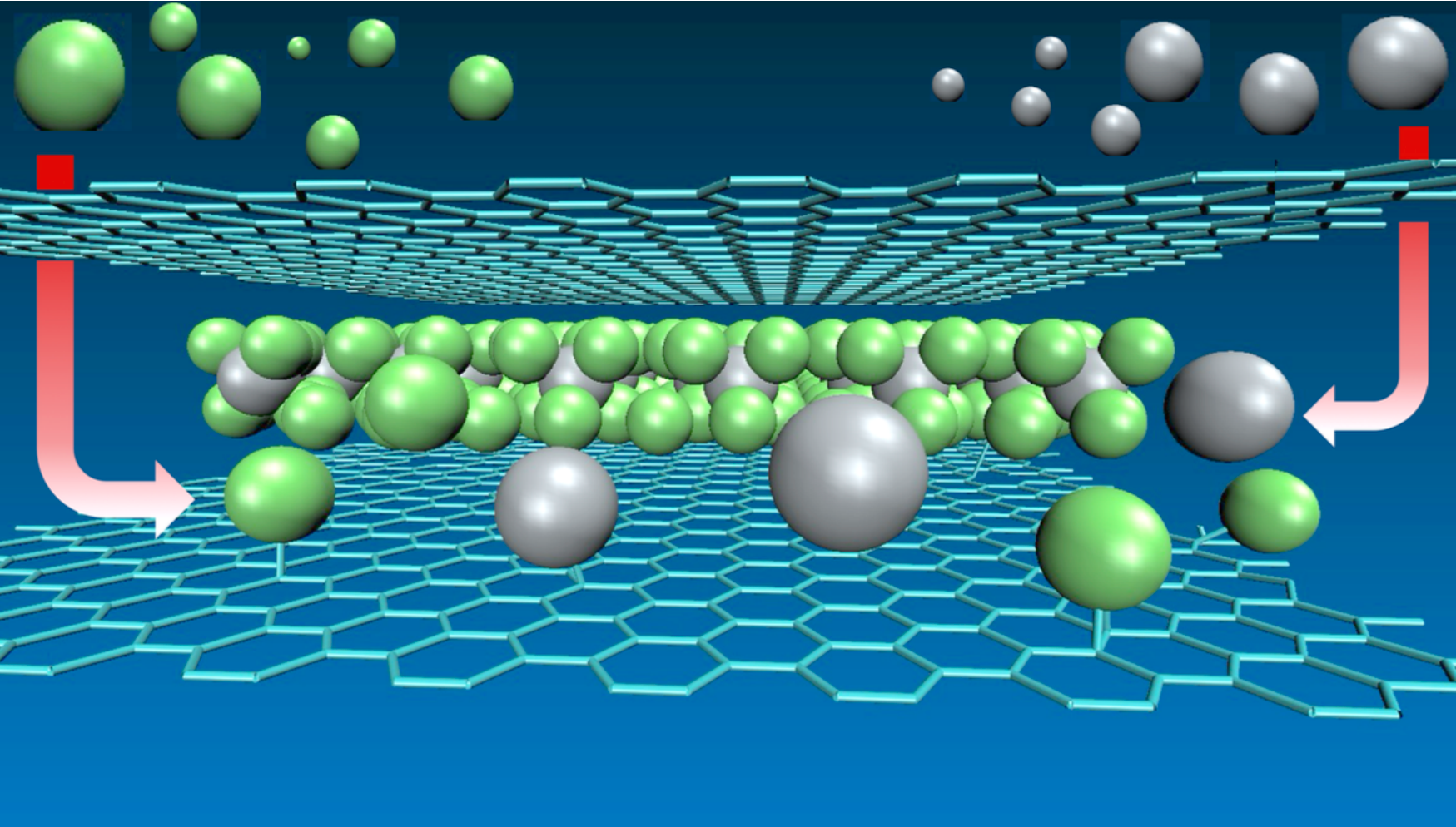}
    \captionsetup{labelformat=empty}
    \addtocounter{figure}{-1}
    \caption{TOC figure}
\end{figure}


\vspace{2pc}
%
%\begin{tocentry}
%\includegraphics[width=8.0cm]{TOC.png}
%\vspace{1cm}
%\\
%
%\end{tocentry}

%\section{Introduction}
%\vspace{1cm} 
%\begin{Block} 

Intercalation of various atom and molecular species into layered
materials has recently been at the forefront of research in materials science  as an important
phenomenon directly relevant to energy/ion storage~\cite{xu2017recent,zhang2021energy} and tuning the
electronic properties of the materials~\cite{Wan2016,Rajapakse2021}. 
In particular, intercalation and
de-intercalation of alkali metal (AM) atoms into graphitic carbon systems\cite{sonia2017},
inorganic layered materials~\cite{zhang2021energy} and their heterostructures~\cite{LIU2020470}
has been
extensively studied in the context of anode operation in electric
batteries. Intercalation of AM atoms into transition metal dichalcogenides (TMDs)
can be also used to induce phase transformations in these systems\cite{Gao2015}. The intercalation
of organic and inorganic molecules into TMDs and
other van der Waals (vdW) materials has been demonstrated to be a powerful
tool for tuning material properties, e.g., adding
superconductivity and magnetism, see Refs. \cite{Wan2016,Rajapakse2021} for an overview.

The isolation of graphene sheets and other two-dimensional materials (2DMs) followed by the
development of vdW heterostructures opened new directions in this research. The chemical
inertness and mechanical robustness of graphene  allowed one to
use graphene as a substrate to grow and stabilize essentially free-standing
2DMs. Moreover, in addition to 2DMs which have layered bulk counterparts, such as 
TMDs \cite{Eichfeld2015,VanEfferen2024},
and can be exfoliated by various techniques,
insulating 2D silica~\cite{Huang2012}, semiconducting 
PdI$_2$\cite{Sinha2020}, or 
metallic AuCu \cite{Zagler2020}, which are either not layered in the bulk form~\cite{Huang2012,Zagler2020} or are difficult to exfoliate\cite{Sinha2020}, were manufactured.
Moreover, as graphene is stable under electron beam at electron energies below 80 keV~\cite{Meyer2012prl}, the  
direct information on the atomic structure can be obtained  using high-resolution transmission electron
microscopy (HR-TEM) due to the graphene's ‘transparency’ to the electrons~\cite{Zhang2019golb}. 

2DMs~\cite{Cao2015nl,Zan2013_ACSNano,Algara-Siller2013_APL,Nguyen2017,Lehnert2021},
liquids~\cite{Textor18}, and 
and soft materials like DNA strands~\cite{Chen13nl}  on graphene
can further be stabilized by placing another graphene sheet on top.
It was also discovered that the protecting action of graphene can not only
prevent marginally stable 2D materials from decomposition in air,
%~\cite{Algara-Siller2013_APL,Zan2013_ACSNano}, 
but together with the electron beam facilitate  phase transformations in the
materials thus obtaining completely new 2D systems~\cite{Lin2021advm,Liu2023acs,Koster2024}.
Moreover, direct intercalation of atomic or molecular species~\cite{Lin2021advm,Liu2023acs,Bonacum2019} 
or ion implantation~\cite{Langle2024} into bilayer graphene (BLG)
can be done. This can give rise to formation of new phases, as apparently
more material can be intercalated into BLG than between graphene sheets in graphite, as
it is much easier to 'push' two free-standing layers away, than those in the bulk layered counterpart. 
At the same time, pressure up to one GPa can be exerted on the encapsulated 2DMs~\cite{Vasu2016}, 
especially close to the material edges in graphene pockets, which, together with charge transfer to/from graphene,
can affect material structure and properties.

The effects of confinement and also higher flexibility (that is the
ability to provide more space) of bilayer graphene as compared to
graphite gave rise to unexpected behavior of intercalants. For example,
formation of multi-layer AM structures has been observed in
BLG~\cite{kuhne2018reversible,Lin2024nc}, contrary to a wide-spread belief that alkali
metal atoms intercalated into layered materials form single-layer
structures only.
Exotic 2DMs can also be created by intercalation of atoms under graphene on 
metal substrates ~\cite{Petrovic2013} or under graphene on monocrystalline SiC 
surface\cite{AlBalushi2016,Pecz2021,Forti2020}, but the materials are normally not free-standing, but covalently 
bonded to the substrate in this
case, and besides direct TEM characterization of the material is not possible. On the contrary,
after intercalation into BLG on a TEM grid, the atomic structure of the system can be
thoroughly investigated. Many other materials can be potentially grown and characterized using this approach.

In this mini-review we summarize the
progress in the material development inside BLG and outline the challenges both experiment and theory
are facing at the moment. Our ultimate goal is to attract the attention of
the scientific community to this relatively simple but efficient
approach to make new 2DMs inside BLG and potentially other bilayers by utilizing 
the stabilization effects of the protective layers and also tune their properties by
charge transfer, pressure built-up related to spatial confinement, and electron beam irradiation. 
We do not focus on the synthesis details, as the methods are very close to the
standard techniques used for the intercalation of various species  into layered materials,
see Refs. \cite{Wan2016,Rajapakse2021,Stark2019}

We start by briefly outlining the
approaches which can be employed to create 2D materials confined inside BLG, then we dwell upon
the  materials which have been manufactured. Finally, we discuss unique opportunities offered by the intercalation into
2DMs beyond graphene and their heterostructures.

%%%%%%%%%%%%%%%%%%%%%%%%

\begin{figure}[!ht] \centering
\includegraphics[width=0.99\linewidth]{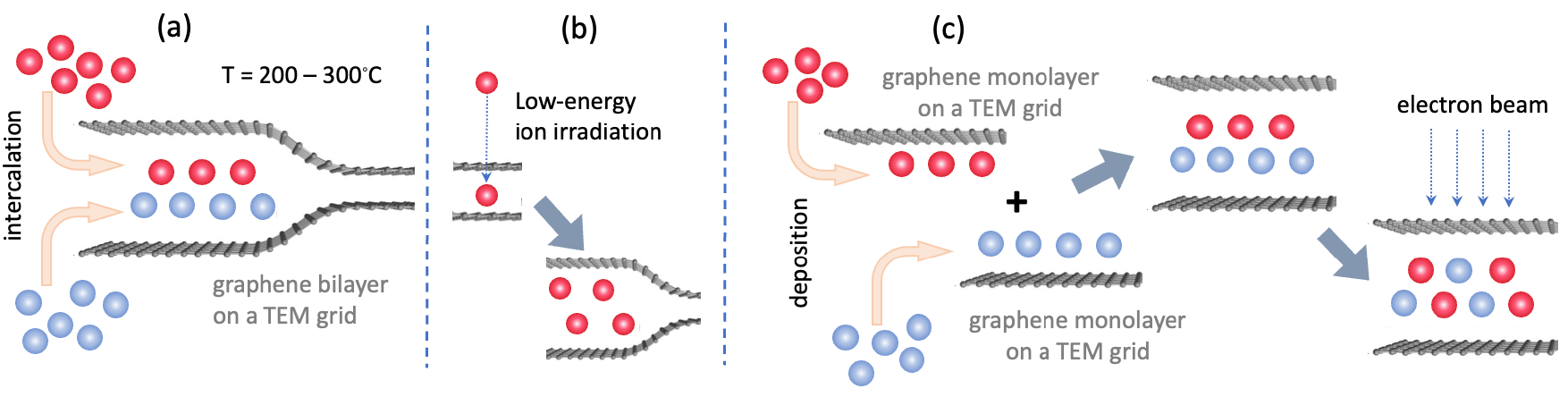}
\caption{\small{Schematic representation of the approaches which can be used
to manufacture new 2D materials or unusual spatially-confined phases of known materials in BLG
on a TEM grid.
(a) Intercalation of atoms and molecules into BLG at elevated temperatures. (b)
Direct low-energy ion implantation into BLG possibly combined with mild annealing.
(c) Deposition of materials on single-layer graphene on TEM grids, followed
by making a 'sandwich'. The material can be further transformed due to exposure to the
electron beam in a TEM, as in case of direct intercalation.
}} 
\label{fig:schematic} 
\end{figure}

Several approaches can be used to develop new materials or unusual 
spatially-confined phases of known materials in BLG. Fig. \ref{fig:schematic} schematically  presents
 the approaches. The atomic or molecular species can directly be
intercalated into BLG at room or elevated temperatures, Fig. \ref{fig:schematic}(a), followed by 
chemical reactions in the confined space, which can also be stimulated by the electron beam.

Low-energy ion implantation into BLG can also be employed, Fig. \ref{fig:schematic}(b).
%as recently demonstrated ~\cite{Langle2024}. 
The choice of the ion energy is obviously of utmost importance,
as ions would go through the system when the energy is too high, and would bounce back when the
energy is too low.  

Another approach is the deposition of materials on single-layer graphene on TEM grids, followed
by making a 'sandwich',
%~\cite{XXX}, 
as illustrated in Fig. \ref{fig:schematic}(c). 
The structure of the material can be further transformed due to exposure to the
electron beam in a TEM, as in case of direct intercalation.

Finally, AM atoms can be driven into BLG by applying voltage to the systems. The setup 
involves BLG either on a substrate or a TEM grid, electrolyte and AM ions, as discussed below.

\begin{figure}[!ht] \centering
\includegraphics[width=0.99\linewidth]{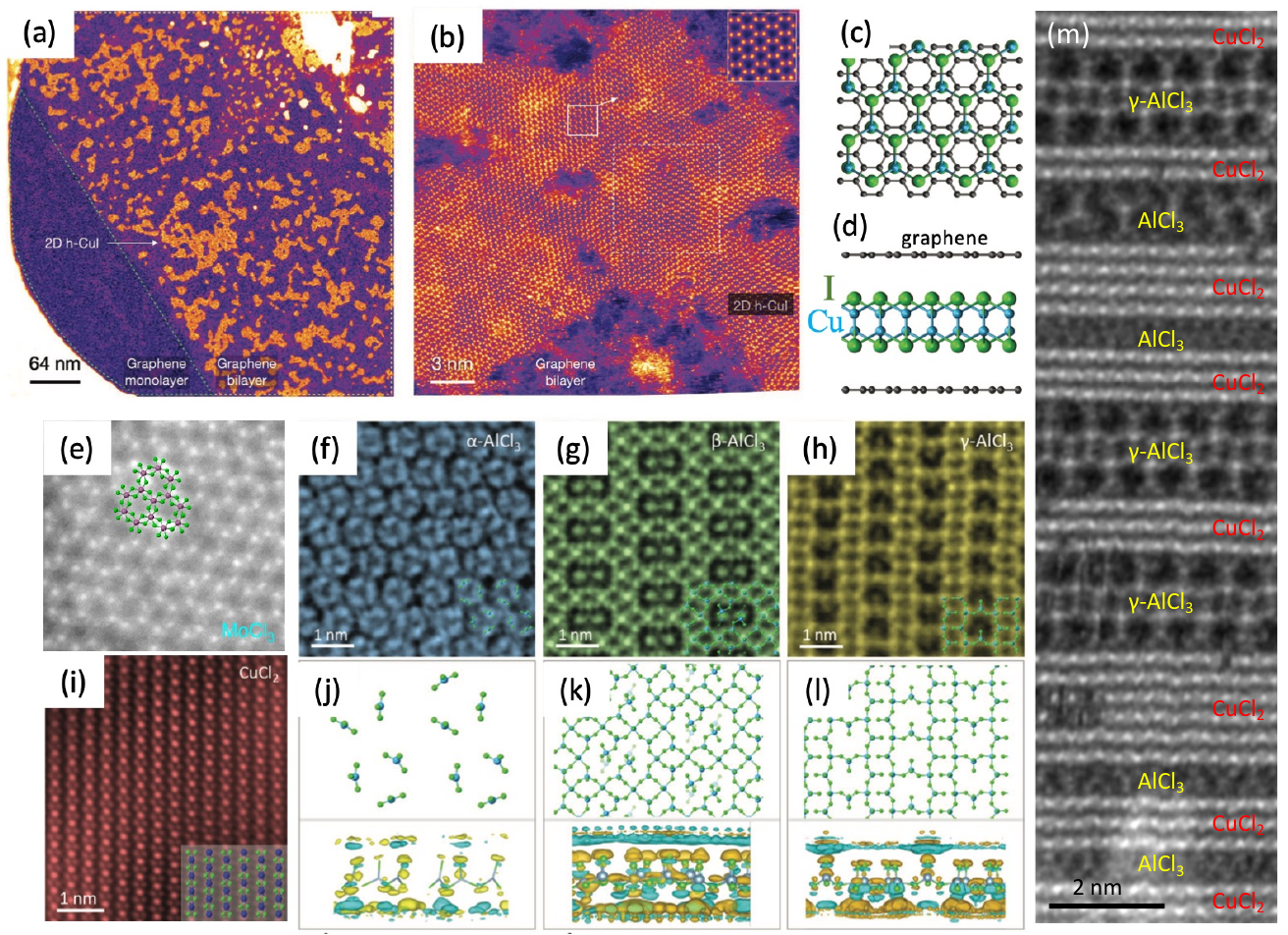}
\caption{\small{
Inorganic 2D materials between graphene sheets of BLG.
(a) TEM image of monolayer h-CuI crystals encapsulated in BLG. 
Note that no h-CuI is visible on the monolayer area on the left-hand side. 
(b) Atomically resolved TEM image of a single 2D h-CuI crystal with a magnifying inset 
in the top right corner. (c-d) Top and side views of the atomic structure of h-CuI, as 
revealed by DFT calculations. Reprinted with permission from Ref. \cite{Mustonen2022}.
(e-i) 2D MCl$_x$ structures (M = Mo, Al, Cu) in BLG. (j-l) Atomic structure
and charge transfer for AlCl$_3$ phases obtained using first-principles calculations.
(m) AlCl$_3$/CuCl$_2$ heterostructures.
Panel (e) reprinted with permission from Ref. \cite{Liu2023acs} and panels (f-m) from 
Ref. \cite{Lin2021advm}.
}} 
\label{fig:phases} 
\end{figure}

Having introduced the methods which can be used to manufacture materials inside BLG, we move on 
the specific materials produced using the approaches discussed above. The results obtained so far are
also summarized in Table 1.

{\bf Covalently bonded inorganic 2D materials}. Single layers of CuI, a material that normally only occurs in the 
layered form at high temperatures between 645 and 675 K, were manufactured by a single-step
wet chemical process involving liquid phase filtration of graphene oxide combined with metal 
chloride solution and hydroiodic acid~\cite{Mustonen2022}. Graphene oxide  was eventually 
converted to BLG, and the materials encapsulated
between graphene sheets was stable at  ambient conditions.
This new 2DM was named by the authors
hexagonal copper iodide (2D h-CuI).
A TEM image of monolayer h-CuI crystals encapsulated in BLG is shown in Fig.~\ref{fig:phases}(a). 
Note that no h-CuI was observed in the monolayer graphene area, which emphasizes the importance
of encapsulation. Atomically resolved TEM image of a single 2D h-CuI crystal with a magnifying inset 
in the top right corner is presented in Fig.~\ref{fig:phases}(b).
The material was characterized by various techniques, including
a combination of atomic-resolution scanning TEM (STEM) 
and ptychographic imaging, electron diffraction, X-ray
absorption spectroscopy, spatially resolved electron energy loss
spectroscopy (EELS). Together with the results of density functional theory (DFT)
calculations, Fig.~\ref{fig:phases}(c,d), the experimental data provided insights into
the atomic structure of the encapsulated CuI
system. Moreover, using the same approach 2D silver iodide (AgI) and nickel iodide (NiI$_2$) 
crystals were manufactured~\cite{Mustonen2022}, thus pointing out that other exotic encapsulated 2D materials
can also be grown. 

Various phases of 2D metal chlorides with different stoichiometries
were produced in BLG~\cite{Lin2021advm,Liu2023acs,Bonacum2019} by intercalation. % FeCl -- to be published?
Specifically, MoCl$_x$ sheets were intercalated into BLG on TEM grids
by placing the grid and MoCl$_5$ powder into glass tube and heating it
to 250$^\circ$C for 12 h in a furnace~\cite{Liu2023acs}. During the heating the MoCl$_5$ powder turned into gas
and intercalated into the area between the graphene
layers. The prepared samples were opened in a glovebox and transferred to the TEM
with minimal exposure to the atmosphere to prevent oxidation.
TEM characterization revealed that  the
intercalated material represents MoCl$_3$ networks, Fig. \ref{fig:phases}(e), MoCl$_2$ chains, and
Mo$_5$Cl$_{10}$ rings. Exposure to the electron beam and possibly charge transfer from graphene 
gave rise to giant lattice distortions 
and frequent structural
transformations, which have never been observed in metal
chloride systems. 

Iron chloride molecules were intercalated into BLG~\cite{Bonacum2019}.
Two distinct intercalated 2D systems were found using TEM/STEM and Raman spectroscopy. 
They were identified as monolayer FeCl$_3$ and  FeCl$_2$. 
These structures are magnetic, which may offer a way to study magnetism in a system with the reduced dimensionality.
It was also found that the
electron beam can convert the FeCl$_3$  monolayer into FeOCl
monolayers with a rectangular lattice. 

Other unprecedented 2D metal chloride structures were produced in BLG
through intercalation of metal and chlorine atoms via chemical vapor transport
inside a vacuum-sealed glass tube~\cite{Lin2021advm}.
In particular, several spatially confined 2D phases of AlCl$_3$ distinct
from their typical bulk forms were observed using HR-TEM, along with the transformations
between the phases likely induced by the electron beam.
Figs. \ref{fig:phases}(f-h) show the STEM images
of the phases of 2D AlCl$_3$, while Figs. \ref{fig:phases}(j-l) present the  atomic structure
and charge transfer for AlCl$_3$ as obtained using first-principles calculations,
which confirmed the metastability of the atomic
structures derived from the experimental images. They also provided insights into
the electronic properties of the phases: They were found to range from insulators to
semimetals. 2D CuCl$_2$ sheets (which in fact are quasi-one-dimensional chains parallel to each other) 
were also synthesized, Fig.~\ref{fig:phases}(i).

Additionally, completely new hybrid systems were produced by co-intercalation of different metal
chlorides. Specifically, in-plane
AlCl$_3$/CuCl$_2$ heterostructures were manufactured, Fig. \ref{fig:phases}(m). The existence of
polymorphic phases and their appearance during exposure to the electron beam
stresses the important role electron irradiation plays. It also hints at
unique possibilities for fabricating new
types of 2D materials with diverse electronic properties confined
between graphene sheets. It should be pointed out, however, that the driving force
behind the transformations between the phases remains unclear.

\begin{figure}[!ht] \centering
\includegraphics[width=0.99\linewidth]{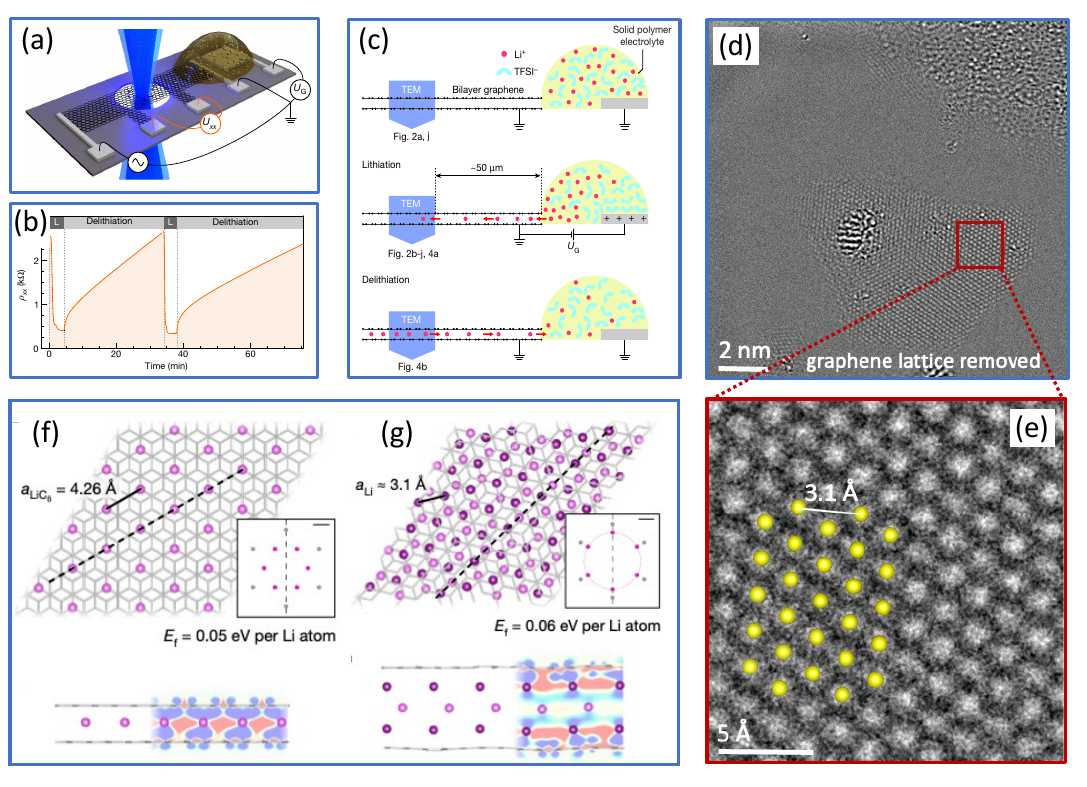}
\caption{\small{Li structures in BLG.
(a) Schematic of the device used in the in-situ TEM experiments where multi-layer
Li structures were formed inside BLG.(b) Bilayer graphene resistivity measured in situ
during two lithiation/delithiation cycles inside the TEM, schematically illustrated in panel (c). 
(d) Atomically-resolved TEM image of a triangular Li crystal with graphene lattice removed. 
(e) Magnified view of the boxed area in (d). The positions of atoms are consistent with the
multi-layer Li fcc structure, but not with the $\sqrt{3} \times \sqrt{3}R30^\circ$ lattice.
Panels (a-c, f) are preprinted with permission from Ref. \cite{kuhne2018reversible}, (d,e) from 
Ref. \cite{YueliangLi24}.
}} 
\label{fig:Li} 
\end{figure}

{\bf 2D metals}.
Intercalation of Li and other AM atoms into BLG has recently been studied both
theoretically\cite{yang2016sodium,lee2012li,chepkasov2020alkali,sonia2017} and 
experimentally~\cite{kuhne2018reversible,sonia2017,ji2019lithium,kuhne2017,Lin2024nc}.
specifically, a macroscopic 
three-dimensional BLG foam was recently manufactured \cite{ji2019lithium}, and its Li-storage capacity and intercalation 
kinetics were investigated. The results indicated that  
Li  atoms can be stored only between the graphene sheets in BLG, not on the outer surfaces.
At the same time, no microscopic data on the arrangement of Li atoms was obtained, but it was
assumed, following other works\cite{yang2016sodium,lee2012li,chepkasov2020alkali,sonia2017}
on Li intercalation into BLG and graphite that Li atoms between graphene sheet form a
$\sqrt{3} \times \sqrt{3}R30^\circ$ lattice, as illustrated in Fig. \ref{fig:Li}(f).

An unexpected Li structures were observed in an {\it in-situ} TEM study~\cite{kuhne2018reversible}, 
which were claimed to be close-packed Li multi-layers between graphene sheets.
Fig. \ref{fig:Li}(a) schematically shows
the setup used in the experiment:  BLG was deposited on a substrate 
with a hole at the centre of the chip and contacted by several
metallic electrodes, that made it possible to measure the resistivity
of the device which is different with and without  intercalated Li atoms, Fig. \ref{fig:Li}(b).
The Li-ion (red spheres)
conducting electrolyte (yellow) connected the BLG to a metallic
counter electrode to form an electrochemical cell.
The hole in the substrate allowed for  in-situ TEM observations of the 
lithiation and delithiation processes. 
Surprisingly, a dense network of Li atoms was found, Fig. \ref{fig:Li}(e), which was not consistent with the
$\sqrt{3} \times \sqrt{3}R30^\circ$ arrangement of atoms,  Fig. \ref{fig:Li}(f).
The observations could be explained through the formation of close-packed multi-layer structures,
as shown in Fig. \ref{fig:Li}(g).
Follow-up studies\cite{YueliangLi24} indicated that Li in BLG mostly forms fcc lattice,
consisting of up to ten layers.  Simultaneous appearance of many triangular Li clusters
was observed. An example is shown in Fig. \ref{fig:Li}(d). This was explained 
through the nucleation of Li structures on 
vacancies in graphene, which have a higher probability to appear under the electron beam
when Li atoms are present~\cite{Zhang2023Li}. The quasi-2D crystallites were found to be oriented in such a way 
that the (111) surface is facing graphene. This is surprising, as the (100) surface was shown to  have
the lowest formation energy in fcc Li crystals\cite{Liu2014Yak}. However, the interface
energy with graphene is the lowest for the (111) surface~\cite{Zhang2023Li}, which made
such orientation energetically favorable. 

The delithiation process was studied as well\cite{YueliangLi24}, and it was found that
upon delithiation the impurity oxygen atoms initially embedded at octahedral interstitial 
positions inside the lithium crystals~\cite{Zhang2023Li} agglomerate at the edges of the crystals, 
thus giving rise to the formation of amorphous dendritic lithium oxide patches, where lithium ions are trapped.
% ARK:
We note that  unambiguous discrimination between the intercalated structures and those on the outer surface 
of BLG in the top view observation still remains a challenge, along with the full characterization
of the atomic structure of the compounds containing foreign atoms (oxygen, hydrogen, etc.). Further
experiments involving EELS chemical analysis are required to differentiate between Li crystals and
oxides or other compounds.

The existence of multi-layers of other AM atoms in BLG,  MoS$_2$ bilayers and their
heterostructures was theoretically
predicted \cite{chepkasov2020alkali,chepkasov2022single}. Although Li forms
covalent bonds with graphitic carbon, Fig. \ref{fig:AMstructures}(j), while for other AMs the interaction is dominated
by charge transfer~\cite{chepkasov2020alkali,moriwake2017sodium,liu2016origin}
between graphene and the AM atoms, multi-layer structures for K are energetically favorable over
single layers, and for Na, Rb, and Cs the energies are close.

Indeed double layers of K, Rb, and Cs inside BLG were observed~\cite{Lin2024nc} later on.
The intercalation of AM atoms into BLG on a TEM grid was done 
through a vapor phase intercalation process. TEM investigations
revealed that the intercalated atoms form double layers inside BLG
with the hcp stacking, and have a C$_6$M$_2$C$_6$ composition, Fig.  \ref{fig:AMstructures}(a-i).
A negative charge
transferred from  AM structure to graphene layers of approximately
$1 - 1.5 \times 10^{14}e^{-}$cm$^{-2}$ was determined by EELS,
Raman, and electrical transport measurements, in agreement with the results
of first-principles calculations, Fig. \ref{fig:AMstructures}(k). Studies on thicker graphene flakes
showed that the double AM layers were absent in the graphite interior,
primarily dominated by single-layer AM intercalation, which again emphasized
different behavior of intercalants in bilayers and bulk layered materials.

It should be pointed out, though, that the projected separations between the AM atoms were smaller than the
DFT calculations predict~\cite{chepkasov2020alkali}. This can be explained by pressure exerted by graphene sheets 
in the direction parallel to the sheets, which can occur when spacing between
the sheets is only partially filled with AM atoms, as illustrated in 
Fig. \ref{fig:AMstructures}(e). This conjecture still remains to be confirmed by
the calculations and/or further experiments. 

Using the electro-chemical deposition of Pd  between graphene
oxide  sheets the growth of few-nm-thick Pd structures was achieved~\cite{Su2019}. The growth was 
self-limiting, which was  a
consequence of the strong interaction of Pd with the confining
sheets, making the growth of bulk  Pd energetically unfavorable. 
Liquid exfoliation of Pd sheets was then demonstrated along with a their high efficiency in  
catalysis and electrocatalysis.

{\bf Noble gas 2D structures.} 2D few-atom noble gas clusters were recently created~\cite{Langle2024} 
between the layers in BLG
using a very interesting approach: low-energy ion irradiation, as schematically 
illustrated in Fig. \ref{fig:schematic}(b). Specifically, Xe and Kr singly-charged
ions  with ultra-low energies about 60 eV were
implanted between suspended graphene sheets. Atomic-resolution characterization
using STEM showed that graphene sheets remain mostly intact so that 
the implanted ions stay between graphene sheets and form
clusters with geometries different from those of free-standing ones. The clusters
were found to be flat due to the pressure of about 0.3 GPa coming from the graphene sheets. 
Their dynamics and atomic transformations at different temperatures were then investigated.  
  
The successful creation of noble gas atom clusters by direct ion implantation indicates 
that this approach can possibly  be used for the encapsulation of other atomic species. The strategies
to 'repair' graphene network should also be developed, by, e.g. adding hydrocarbon molecules.
Moreover, as graphene is robust under electron beam (at electron energies below 80 keV~\cite{Meyer2012prl}),
exposure to the electron irradiation may stimulate chemical reactions and phase
transformation in the confined area of BLG.

\begin{figure}[!ht] \centering
\includegraphics[width=0.99\linewidth]{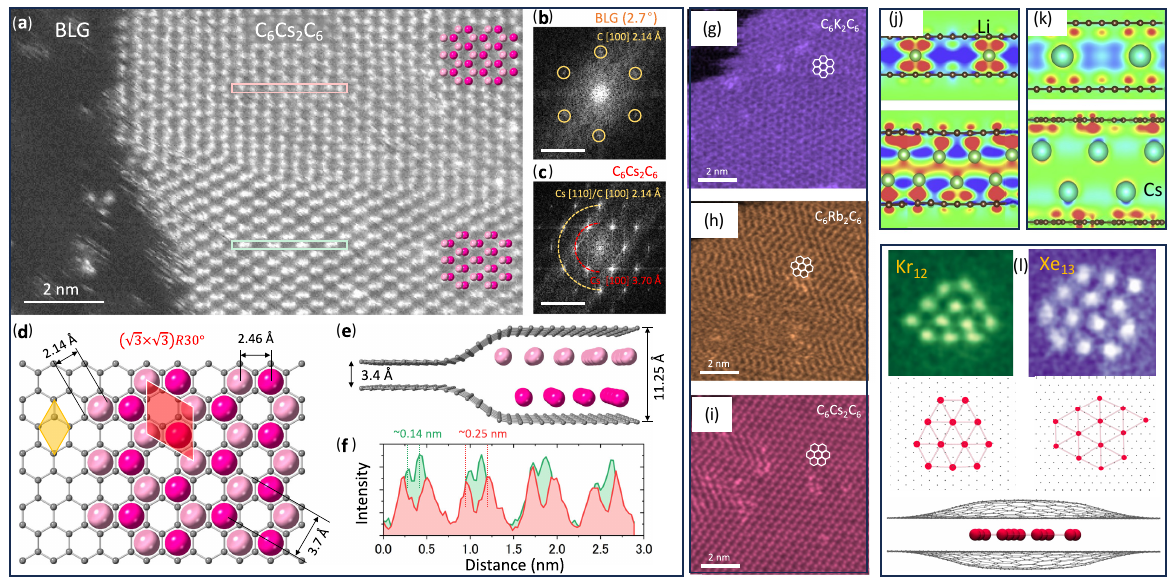}
\caption{
\small{
Unusual alkali metal and inert gas atom systems confined in BLG. (a) STEM image of 
Cs-intercalated BLG
displaying the C$_6$Cs$_2$C$_6$ structure. (b) and (c) Fast Fourier
transform %(FFT) 
image of BLG and Cs domains in (a), scale bar = 5nm$^{-1}$.
%The [100] and [110] spots of C$_6$Cs$_2$C$_6$ domain
%correspond to interplanar d-spacings of 3.70 Å and 2.14 Å in real space,
%with the Cs [110] spots overlapping with the gra- phene [100] spots. 
(d) Top-view atomic model of C$_6$Cs$_2$C$_6$. The yellow rhombus highlights a
graphene (1 × 1) unit cell with $a = b = 2.46$ Å, while the red rhombus
highlights the unit cell of the Cs lattice with $\sqrt{3} \times \sqrt{3}R30^\circ$ lattice, where
a = b = 4.26 Å. (e) Side  view of the C$_6$Cs$_2$C$_6$ structure. 
%The Cs atoms in different atomic planes are color-coded with two different shades ofred. 
(f) ADF profile of the hexagonal Cs layer along the pink and light
green boxes in (a). The red profile displays the shortest distance
between two Cs atoms as 0.25 nm, consistent with the atomic model shown
in (d). The green profile shows the shorter distance between the Cs
atoms, indicating the lateral displacement of the two Cs atomic plans.
(g-i) TEM images of the similar K, Rb, and Cs structures. 
Panels (a-i) reprinted with permission from Ref. \cite{Lin2024nc}.
Charge difference (cross-section through Li (j) and Cs (k) atoms perpendicular to the BLG planes for a single and double layer atoms structures. 
Note a build-up of the electron density between Li atoms and graphene, illustrating a substantial contribution to the bonding from
covalent interaction. Red color corresponds to density build-up, blue to depletion. From Ref. \cite{chepkasov2020alkali}.
(l) Flat Kr and Xe clusters in BLG~\cite{Langle2024} created by ion implantation. 
}}
\label{fig:AMstructures} 
\end{figure}

As evident from a brief summary of the results obtained so far, BLG is a unique platform
to create novel 2DMs by intercalations, as it in comparison to graphite allows for easier intercalation and
a much larger increase in the inter-layer separation of the sheets.  
Potentially other 2DMs, such 
as h-BN or TMDs and their heterostructures can be used. Another interesting direction is to 
create BLG with the controlled twist angle, which affects the behavior of the intercalated material~\cite{Araki2022acs}.
BLG, which is robust, but transparent to the electron in the TEM, makes it possible to get
direct microscopic information on the atomic structure of the encapsulated 2DMs. Moreover,
new phases can appear due to the electron irradiation, and the transformations
can be followed in situ. 

At the same time, many aspects of the intercalation, interaction
of the encapsulated 2DMs with the protecting sheets and effects of the electron beam still lack
complete understanding.
Specifically, the role of pressure and charge transfer should be fully clarified, along with
the actual driving force for the observed phase transformations, which may be related to formation and
dynamics of defects, as in the case of AlCl$_3$\cite{Lin2021advm}.
We note that the lateral sizes of the structures were rather small, so that 
their characterization by larger scale methods was difficult, and one of the
challenges is to increase the lateral size of the encapsulated structures.
As for using ion implantation into BLG to create new materials, it is interesting to explore
if this strategy can be extended from inert gas ions to other chemical elements, find out what
the optimum ion energies could be, and how inevitable damage to graphene sheets can be repaired.

Further developments in this field should enhance our understanding of the behavior of
matter in the confined space and suggest new ways for creating unprecedented 2D systems protected 
by graphene sheets or other robust 2DMs. As some structures, e.g. 2D iron chloride, are magnetic,
this would allow studies on the magnetism in 2D systems, specifically in the context
of information storage. Unique electronic properties and co-existence of different  
structures can allow for addressing a wide range of phenomena, such as Klein 
tunneling of Dirac-like fermions or half-metal behavior. Controlled creation of defects 
in these systems, e.g, using electron beam, can open new avenues for the development
of single-photon emitters, provided that insulating materials, e.g., h-BN is used for encapsulation.
Overall, creation of 2D materials encapsulated into BLG should 
offer a unique opportunity to tune the properties of the system
for specific applications such as photonics,  quantum information and energy storage.

\begin{figure}[!ht] \centering
\includegraphics[width=0.99\linewidth]{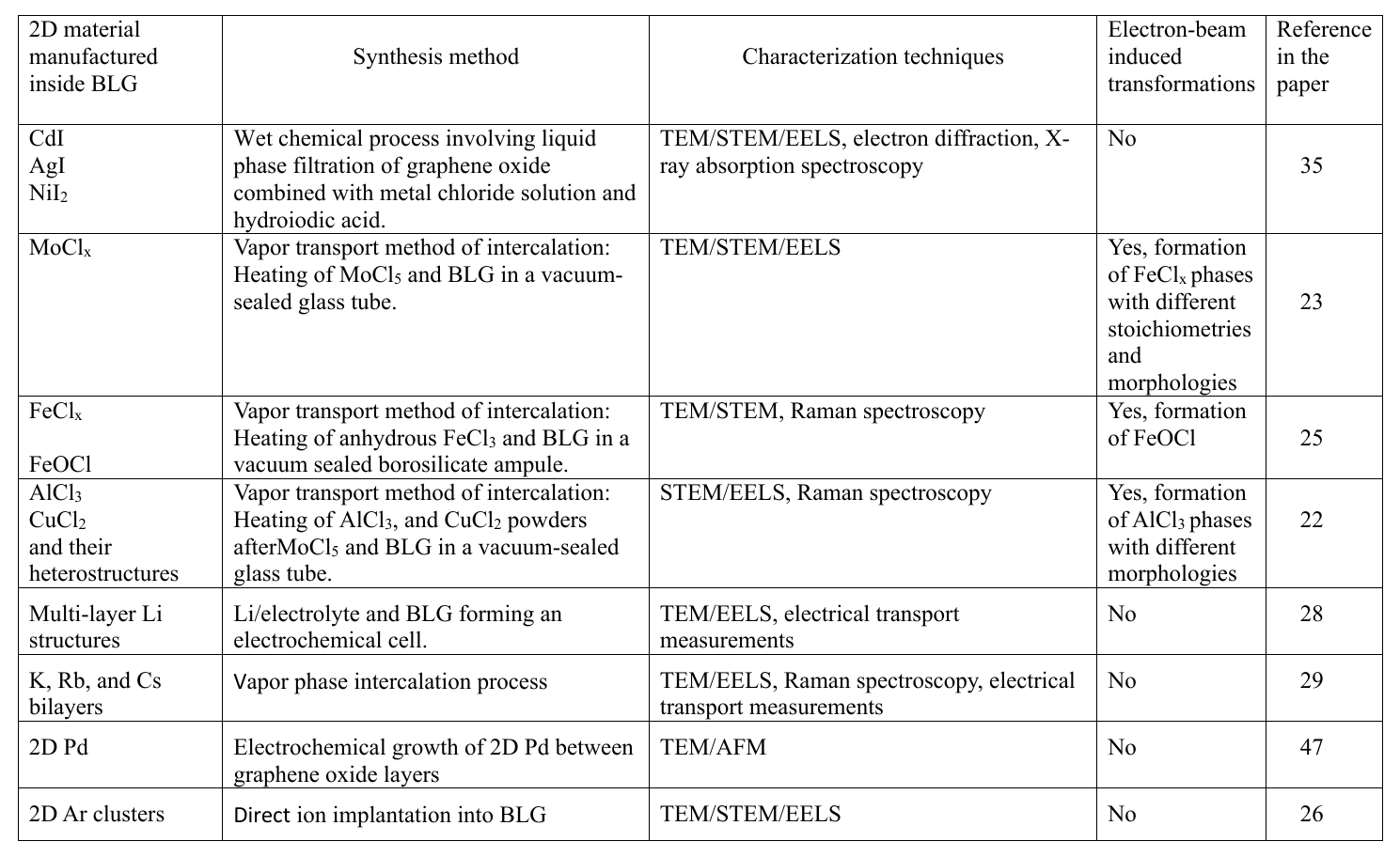}
    \captionsetup{labelformat=empty}
\caption{\small{Table 1. Summary of the encapsulated 2D materials  with unusual structures obtained so far.
}} 
    \addtocounter{figure}{-1}
\label{fig:table} 
\end{figure}
 
\textbf{Acknowledgments.}
We thank U. Kaiser, H. Ago, J. Smet, M. Ghorbani-Asl, S. Kretschmer for many years of collaboration
on the topic of this review. We further thank J. Kotakoski, M. Längle, and H. Åhlgren
for discussions and providing STEM images and atomic models of noble gas atom clusters
between graphene sheets.
AVK acknowledges funding from the German Research Foundation (DFG), 
Project KR 4866/9-1 (406129719) and the collaborative research center “Chemistry of 
Synthetic 2D Materials” SFB- 1415-417590517.
K.S. and Y.C.L. acknowledge to JSPS-KAKENHI (16H06333, 21H05235, 22F22358, 22H05478), 
the JST-CREST program (JPMJCR20B1, JMJCR20B5, JPMJCR1993), ERC “MORE-TEM” (951215), 
and the JSPS A3 Foresight Program.  

\bibliography{nlrev}

\providecommand{\latin}[1]{#1}
\makeatletter
\providecommand{\doi}
  {\begingroup\let\do\@makeother\dospecials
  \catcode`\{=1 \catcode`\}=2 \doi@aux}
\providecommand{\doi@aux}[1]{\endgroup\texttt{#1}}
\makeatother
\providecommand*\mcitethebibliography{\thebibliography}
\csname @ifundefined\endcsname{endmcitethebibliography}
  {\let\endmcitethebibliography\endthebibliography}{}
\begin{mcitethebibliography}{49}
\providecommand*\natexlab[1]{#1}
\providecommand*\mciteSetBstSublistMode[1]{}
\providecommand*\mciteSetBstMaxWidthForm[2]{}
\providecommand*\mciteBstWouldAddEndPuncttrue
  {\def\EndOfBibitem{\unskip.}}
\providecommand*\mciteBstWouldAddEndPunctfalse
  {\let\EndOfBibitem\relax}
\providecommand*\mciteSetBstMidEndSepPunct[3]{}
\providecommand*\mciteSetBstSublistLabelBeginEnd[3]{}
\providecommand*\EndOfBibitem{}
\mciteSetBstSublistMode{f}
\mciteSetBstMaxWidthForm{subitem}{(\alph{mcitesubitemcount})}
\mciteSetBstSublistLabelBeginEnd
  {\mcitemaxwidthsubitemform\space}
  {\relax}
  {\relax}

\bibitem[Xu \latin{et~al.}(2017)Xu, Dou, Wei, Ma, Deng, Li, Liu, and
  Dou]{xu2017recent}
Xu,~J.; Dou,~Y.; Wei,~Z.; Ma,~J.; Deng,~Y.; Li,~Y.; Liu,~H.; Dou,~S. Recent
  Progress in Graphite Intercalation Compounds for Rechargeable Metal (Li, Na,
  K, Al)-Ion Batteries. \emph{Advanced Science} \textbf{2017}, \emph{4},
  1700146\relax
\mciteBstWouldAddEndPuncttrue
\mciteSetBstMidEndSepPunct{\mcitedefaultmidpunct}
{\mcitedefaultendpunct}{\mcitedefaultseppunct}\relax
\EndOfBibitem
\bibitem[Zhang \latin{et~al.}(2021)Zhang, Lai, Wang, Chen, and
  Shen]{zhang2021energy}
Zhang,~J.; Lai,~L.; Wang,~H.; Chen,~M.; Shen,~Z.~X. Energy storage mechanisms
  of anode materials for potassium ion batteries. \emph{Materials Today Energy}
  \textbf{2021}, 100747\relax
\mciteBstWouldAddEndPuncttrue
\mciteSetBstMidEndSepPunct{\mcitedefaultmidpunct}
{\mcitedefaultendpunct}{\mcitedefaultseppunct}\relax
\EndOfBibitem
\bibitem[Wan \latin{et~al.}(2016)Wan, Lacey, Dai, Bao, Fuhrer, and Hu]{Wan2016}
Wan,~J.; Lacey,~S.~D.; Dai,~J.; Bao,~W.; Fuhrer,~M.~S.; Hu,~L. {Tuning
  two-dimensional nanomaterials by intercalation: Materials, properties and
  applications}. \emph{Chemical Society Reviews} \textbf{2016}, \emph{45},
  6742--6765\relax
\mciteBstWouldAddEndPuncttrue
\mciteSetBstMidEndSepPunct{\mcitedefaultmidpunct}
{\mcitedefaultendpunct}{\mcitedefaultseppunct}\relax
\EndOfBibitem
\bibitem[Rajapakse \latin{et~al.}(2021)Rajapakse, Karki, Abu, Pishgar, Musa,
  Riyadh, Yu, Sumanasekera, and Jasinski]{Rajapakse2021}
Rajapakse,~M.; Karki,~B.; Abu,~U.~O.; Pishgar,~S.; Musa,~M. R.~K.; Riyadh,~S.
  M.~S.; Yu,~M.; Sumanasekera,~G.; Jasinski,~J.~B. {Intercalation as a
  versatile tool for fabrication, property tuning, and phase transitions in 2D
  materials}. \emph{npj 2D Materials and Applications} \textbf{2021}, \emph{5},
  30\relax
\mciteBstWouldAddEndPuncttrue
\mciteSetBstMidEndSepPunct{\mcitedefaultmidpunct}
{\mcitedefaultendpunct}{\mcitedefaultseppunct}\relax
\EndOfBibitem
\bibitem[Sonia \latin{et~al.}(2017)Sonia, Jangid, Ananthoju, Aslam, Johari, and
  Mukhopadhyay]{sonia2017}
Sonia,~F.~J.; Jangid,~M.~K.; Ananthoju,~B.; Aslam,~M.; Johari,~P.;
  Mukhopadhyay,~A. Understanding the Li-storage in few layers graphene with
  respect to bulk graphite: experimental, analytical and computational study.
  \emph{Journal of Materials Chemistry A} \textbf{2017}, \emph{5},
  8662--8679\relax
\mciteBstWouldAddEndPuncttrue
\mciteSetBstMidEndSepPunct{\mcitedefaultmidpunct}
{\mcitedefaultendpunct}{\mcitedefaultseppunct}\relax
\EndOfBibitem
\bibitem[Liu \latin{et~al.}(2020)Liu, Bai, Zhao, Yao, and Pang]{LIU2020470}
Liu,~C.; Bai,~Y.; Zhao,~Y.; Yao,~H.; Pang,~H. MoS2/graphene composites:
  Fabrication and electrochemical energy storage. \emph{Energy Storage
  Materials} \textbf{2020}, \emph{33}, 470--502\relax
\mciteBstWouldAddEndPuncttrue
\mciteSetBstMidEndSepPunct{\mcitedefaultmidpunct}
{\mcitedefaultendpunct}{\mcitedefaultseppunct}\relax
\EndOfBibitem
\bibitem[Gao \latin{et~al.}(2015)Gao, Wang, Zhang, Huang, and Liu]{Gao2015}
Gao,~P.; Wang,~L.; Zhang,~Y.; Huang,~Y.; Liu,~K. {Atomic-Scale Probing of the
  Dynamics of Sodium Transport and Intercalation-Induced Phase Transformations
  in MoS2}. \emph{ACS Nano} \textbf{2015}, \emph{9}, 11296--11301\relax
\mciteBstWouldAddEndPuncttrue
\mciteSetBstMidEndSepPunct{\mcitedefaultmidpunct}
{\mcitedefaultendpunct}{\mcitedefaultseppunct}\relax
\EndOfBibitem
\bibitem[Eichfeld \latin{et~al.}(2015)Eichfeld, Hossain, Lin, Piasecki, Kupp,
  Birdwell, Burke, Lu, Peng, Li, Azcatl, McDonnell, Wallace, Kim, Mayer,
  Redwing, and Robinson]{Eichfeld2015}
Eichfeld,~S.~M.; Hossain,~L.; Lin,~Y.~C.; Piasecki,~A.~F.; Kupp,~B.;
  Birdwell,~A.~G.; Burke,~R.~A.; Lu,~N.; Peng,~X.; Li,~J.; Azcatl,~A.;
  McDonnell,~S.; Wallace,~R.~M.; Kim,~M.~J.; Mayer,~T.~S.; Redwing,~J.~M.;
  Robinson,~J.~A. {Highly scalable, atomically thin WSe2grown via metal-organic
  chemical vapor deposition}. \emph{ACS Nano} \textbf{2015}, \emph{9},
  2080--2087\relax
\mciteBstWouldAddEndPuncttrue
\mciteSetBstMidEndSepPunct{\mcitedefaultmidpunct}
{\mcitedefaultendpunct}{\mcitedefaultseppunct}\relax
\EndOfBibitem
\bibitem[van Efferen \latin{et~al.}(2024)van Efferen, Hall, Atodiresei, Boix,
  Safeer, Wekking, Vinogradov, Preobrajenski, Knudsen, Fischer, Jolie, and
  Michely]{VanEfferen2024}
van Efferen,~C.; Hall,~J.; Atodiresei,~N.; Boix,~V.; Safeer,~A.; Wekking,~T.;
  Vinogradov,~N.~A.; Preobrajenski,~A.~B.; Knudsen,~J.; Fischer,~J.; Jolie,~W.;
  Michely,~T. {2D Vanadium Sulfides: Synthesis, Atomic Structure Engineering,
  and Charge Density Waves}. \emph{ACS Nano} \textbf{2024}, \emph{18},
  14161--14175\relax
\mciteBstWouldAddEndPuncttrue
\mciteSetBstMidEndSepPunct{\mcitedefaultmidpunct}
{\mcitedefaultendpunct}{\mcitedefaultseppunct}\relax
\EndOfBibitem
\bibitem[Huang \latin{et~al.}(2012)Huang, Kurasch, Srivastava, Skakalova,
  Kotakoski, Krasheninnikov, Hovden, Mao, Meyer, Smet, Muller, and
  Kaiser]{Huang2012}
Huang,~P.~Y.; Kurasch,~S.; Srivastava,~A.; Skakalova,~V.; Kotakoski,~J.;
  Krasheninnikov,~A.~V.; Hovden,~R.; Mao,~Q.; Meyer,~J.~C.; Smet,~J.;
  Muller,~D.~A.; Kaiser,~U. {Direct Imaging of a Two-Dimensional Silica Glass
  on Graphene}. \emph{Nano Letters} \textbf{2012}, \emph{12}, 1081--1086\relax
\mciteBstWouldAddEndPuncttrue
\mciteSetBstMidEndSepPunct{\mcitedefaultmidpunct}
{\mcitedefaultendpunct}{\mcitedefaultseppunct}\relax
\EndOfBibitem
\bibitem[Sinha \latin{et~al.}(2020)Sinha, Zhu, France-Lanord, Sheng, Grossman,
  Porfyrakis, and Warner]{Sinha2020}
Sinha,~S.; Zhu,~T.; France-Lanord,~A.; Sheng,~Y.; Grossman,~J.~C.;
  Porfyrakis,~K.; Warner,~J.~H. {Atomic structure and defect dynamics of
  monolayer lead iodide nanodisks with epitaxial alignment on graphene}.
  \emph{Nature Communications} \textbf{2020}, \emph{11}, 823\relax
\mciteBstWouldAddEndPuncttrue
\mciteSetBstMidEndSepPunct{\mcitedefaultmidpunct}
{\mcitedefaultendpunct}{\mcitedefaultseppunct}\relax
\EndOfBibitem
\bibitem[Zagler \latin{et~al.}(2020)Zagler, Reticcioli, Mangler, Scheinecker,
  Franchini, and Kotakoski]{Zagler2020}
Zagler,~G.; Reticcioli,~M.; Mangler,~C.; Scheinecker,~D.; Franchini,~C.;
  Kotakoski,~J. {CuAu, a hexagonal two-dimensional metal}. \emph{2D Materials}
  \textbf{2020}, \emph{7}, 045017\relax
\mciteBstWouldAddEndPuncttrue
\mciteSetBstMidEndSepPunct{\mcitedefaultmidpunct}
{\mcitedefaultendpunct}{\mcitedefaultseppunct}\relax
\EndOfBibitem
\bibitem[Meyer \latin{et~al.}(2012)Meyer, Eder, Kurasch, Skakalova, Kotakoski,
  Park, Roth, Chuvilin, Eyhusen, Benner, Krasheninnikov, and
  Kaiser]{Meyer2012prl}
Meyer,~J.~C.; Eder,~F.; Kurasch,~S.; Skakalova,~V.; Kotakoski,~J.; Park,~H.~J.;
  Roth,~S.; Chuvilin,~A.; Eyhusen,~S.; Benner,~G.; Krasheninnikov,~A.~V.;
  Kaiser,~U. {Accurate Measurement of Electron Beam Induced Displacement Cross
  Sections for Single-Layer Graphene}. \emph{Physical Review Letters}
  \textbf{2012}, \emph{108}, 196102\relax
\mciteBstWouldAddEndPuncttrue
\mciteSetBstMidEndSepPunct{\mcitedefaultmidpunct}
{\mcitedefaultendpunct}{\mcitedefaultseppunct}\relax
\EndOfBibitem
\bibitem[Zhang \latin{et~al.}(2020)Zhang, Firestein, Fernando, Siriwardena, von
  Treifeldt, and Golberg]{Zhang2019golb}
Zhang,~C.; Firestein,~K.~L.; Fernando,~J. F.~S.; Siriwardena,~D.; von
  Treifeldt,~J.~E.; Golberg,~D. {Recent Progress of In Situ Transmission
  Electron Microscopy for Energy Materials}. \emph{Advanced Materials}
  \textbf{2020}, \emph{32}, 1904094\relax
\mciteBstWouldAddEndPuncttrue
\mciteSetBstMidEndSepPunct{\mcitedefaultmidpunct}
{\mcitedefaultendpunct}{\mcitedefaultseppunct}\relax
\EndOfBibitem
\bibitem[Cao \latin{et~al.}(2015)Cao, Mishchenko, Yu, Khestanova, Rooney,
  Prestat, Kretinin, Blake, Shalom, Woods, Chapman, Balakrishnan, Grigorieva,
  Novoselov, Piot, Potemski, Watanabe, Taniguchi, Haigh, Geim, and
  Gorbachev]{Cao2015nl}
Cao,~Y.; Mishchenko,~A.; Yu,~G.~L.; Khestanova,~E.; Rooney,~a.~P.; Prestat,~E.;
  Kretinin,~a.~V.; Blake,~P.; Shalom,~M.~B.; Woods,~C.; Chapman,~J.;
  Balakrishnan,~G.; Grigorieva,~I.~V.; Novoselov,~K.~S.; Piot,~B.~a.;
  Potemski,~M.; Watanabe,~K.; Taniguchi,~T.; Haigh,~S.~J.; Geim,~A.~K.
  \latin{et~al.}  {Quality Heterostructures from Two-Dimensional Crystals
  Unstable in Air by Their Assembly in Inert Atmosphere}. \emph{Nano Letters}
  \textbf{2015}, \emph{15}, 4914--4921\relax
\mciteBstWouldAddEndPuncttrue
\mciteSetBstMidEndSepPunct{\mcitedefaultmidpunct}
{\mcitedefaultendpunct}{\mcitedefaultseppunct}\relax
\EndOfBibitem
\bibitem[Zan \latin{et~al.}(2013)Zan, Ramasse, Jalil, Georgiou, Bangert, and
  Novoselov]{Zan2013_ACSNano}
Zan,~R.; Ramasse,~Q.~M.; Jalil,~R.; Georgiou,~T.; Bangert,~U.; Novoselov,~K.~S.
  Control of radiation damage in MoS2 by graphene encapsulation. \emph{ACS
  Nano} \textbf{2013}, \emph{7}, 10167--10174\relax
\mciteBstWouldAddEndPuncttrue
\mciteSetBstMidEndSepPunct{\mcitedefaultmidpunct}
{\mcitedefaultendpunct}{\mcitedefaultseppunct}\relax
\EndOfBibitem
\bibitem[Algara-Siller \latin{et~al.}(2013)Algara-Siller, Kurasch, Sedighi,
  Lehtinen, and Kaiser]{Algara-Siller2013_APL}
Algara-Siller,~G.; Kurasch,~S.; Sedighi,~M.; Lehtinen,~O.; Kaiser,~U. The
  pristine atomic structure of MoS2 monolayer protected from electron radiation
  damage by graphene. \emph{Appl. Phys. Lett.} \textbf{2013}, \emph{103},
  203107\relax
\mciteBstWouldAddEndPuncttrue
\mciteSetBstMidEndSepPunct{\mcitedefaultmidpunct}
{\mcitedefaultendpunct}{\mcitedefaultseppunct}\relax
\EndOfBibitem
\bibitem[Nguyen \latin{et~al.}(2017)Nguyen, Komsa, Khestanova, Kashtiban,
  Peters, Lawlor, Sanchez, Sloan, Gorbachev, Grigorieva, Krasheninnikov, and
  Haigh]{Nguyen2017}
Nguyen,~L.; Komsa,~H.-P.; Khestanova,~E.; Kashtiban,~R.~J.; Peters,~J. J.~P.;
  Lawlor,~S.; Sanchez,~A.~M.; Sloan,~J.; Gorbachev,~R.~V.; Grigorieva,~I.~V.;
  Krasheninnikov,~A.~V.; Haigh,~S.~J. {Atomic Defects and Doping of Monolayer
  NbSe 2}. \emph{ACS Nano} \textbf{2017}, \emph{11}, 2894--2904\relax
\mciteBstWouldAddEndPuncttrue
\mciteSetBstMidEndSepPunct{\mcitedefaultmidpunct}
{\mcitedefaultendpunct}{\mcitedefaultseppunct}\relax
\EndOfBibitem
\bibitem[Lehnert \latin{et~al.}(2021)Lehnert, Kretschmer, Br{\"{a}}uer,
  Krasheninnikov, and Kaiser]{Lehnert2021}
Lehnert,~T.; Kretschmer,~S.; Br{\"{a}}uer,~F.; Krasheninnikov,~A.~V.;
  Kaiser,~U. {Quasi-two-dimensional NaCl crystals encapsulated between graphene
  sheets and their decomposition under an electron beam}. \emph{Nanoscale}
  \textbf{2021}, \emph{13}, 19626--19633\relax
\mciteBstWouldAddEndPuncttrue
\mciteSetBstMidEndSepPunct{\mcitedefaultmidpunct}
{\mcitedefaultendpunct}{\mcitedefaultseppunct}\relax
\EndOfBibitem
\bibitem[Textor and de~Jonge(2018)Textor, and de~Jonge]{Textor18}
Textor,~M.; de~Jonge,~N. Strategies for Preparing Graphene Liquid Cells for
  Transmission Electron Microscopy. \emph{Nano Letters} \textbf{2018},
  \emph{18}, 3313--3321\relax
\mciteBstWouldAddEndPuncttrue
\mciteSetBstMidEndSepPunct{\mcitedefaultmidpunct}
{\mcitedefaultendpunct}{\mcitedefaultseppunct}\relax
\EndOfBibitem
\bibitem[Chen \latin{et~al.}(2013)Chen, Smith, Park, Kim, Ho, Rasool, Zettl,
  and Alivisatos]{Chen13nl}
Chen,~Q.; Smith,~J.~M.; Park,~J.; Kim,~K.; Ho,~D.; Rasool,~H.~I.; Zettl,~A.;
  Alivisatos,~A.~P. 3D Motion of DNA-Au Nanoconjugates in Graphene Liquid Cell
  Electron Microscopy. \emph{Nano Letters} \textbf{2013}, \emph{13},
  4556--4561\relax
\mciteBstWouldAddEndPuncttrue
\mciteSetBstMidEndSepPunct{\mcitedefaultmidpunct}
{\mcitedefaultendpunct}{\mcitedefaultseppunct}\relax
\EndOfBibitem
\bibitem[Lin \latin{et~al.}(2021)Lin, Motoyama, Kretschmer, Ghaderzadeh,
  Ghorbani-Asl, Araki, Krasheninnikov, Ago, and Suenaga]{Lin2021advm}
Lin,~Y.; Motoyama,~A.; Kretschmer,~S.; Ghaderzadeh,~S.; Ghorbani-Asl,~M.;
  Araki,~Y.; Krasheninnikov,~A.~V.; Ago,~H.; Suenaga,~K. {Polymorphic Phases of
  Metal Chlorides in the Confined 2D Space of Bilayer Graphene}. \emph{Advanced
  Materials} \textbf{2021}, \emph{33}, 2105898\relax
\mciteBstWouldAddEndPuncttrue
\mciteSetBstMidEndSepPunct{\mcitedefaultmidpunct}
{\mcitedefaultendpunct}{\mcitedefaultseppunct}\relax
\EndOfBibitem
\bibitem[Liu \latin{et~al.}(2023)Liu, Lin, Kretschmer, Ghorbani-Asl,
  Sol{\'{i}}s-Fern{\'{a}}ndez, Siao, Chiu, Ago, Krasheninnikov, and
  Suenaga]{Liu2023acs}
Liu,~Q.; Lin,~Y.-C.; Kretschmer,~S.; Ghorbani-Asl,~M.;
  Sol{\'{i}}s-Fern{\'{a}}ndez,~P.; Siao,~M.-d.; Chiu,~P.-w.; Ago,~H.;
  Krasheninnikov,~A.~V.; Suenaga,~K. {Molybdenum Chloride Nanostructures with
  Giant Lattice Distortions Intercalated into Bilayer Graphene}. \emph{ACS
  Nano} \textbf{2023}, \emph{17}, 23659--23670\relax
\mciteBstWouldAddEndPuncttrue
\mciteSetBstMidEndSepPunct{\mcitedefaultmidpunct}
{\mcitedefaultendpunct}{\mcitedefaultseppunct}\relax
\EndOfBibitem
\bibitem[K{\"{o}}ster \latin{et~al.}(2024)K{\"{o}}ster, Kretschmer, Storm,
  Rasper, Kinyanjui, Krasheninnikov, and Kaiser]{Koster2024}
K{\"{o}}ster,~J.; Kretschmer,~S.; Storm,~A.; Rasper,~F.; Kinyanjui,~M.~K.;
  Krasheninnikov,~A.~V.; Kaiser,~U. {Phase transformations in single-layer MoTe
  2 stimulated by electron irradiation and annealing}. \emph{Nanotechnology}
  \textbf{2024}, \emph{35}, 145301\relax
\mciteBstWouldAddEndPuncttrue
\mciteSetBstMidEndSepPunct{\mcitedefaultmidpunct}
{\mcitedefaultendpunct}{\mcitedefaultseppunct}\relax
\EndOfBibitem
\bibitem[Bonacum \latin{et~al.}(2019)Bonacum, O'Hara, Bao, Ovchinnikov, Zhang,
  Gordeev, Arora, Reich, Idrobo, Haglund, Pantelides, and Bolotin]{Bonacum2019}
Bonacum,~J.~P.; O'Hara,~A.; Bao,~D.-L.; Ovchinnikov,~O.~S.; Zhang,~Y.-F.;
  Gordeev,~G.; Arora,~S.; Reich,~S.; Idrobo,~J.-C.; Haglund,~R.~F.;
  Pantelides,~S.~T.; Bolotin,~K.~I. {Atomic-resolution visualization and doping
  effects of complex structures in intercalated bilayer graphene}.
  \emph{Physical Review Materials} \textbf{2019}, \emph{3}, 064004\relax
\mciteBstWouldAddEndPuncttrue
\mciteSetBstMidEndSepPunct{\mcitedefaultmidpunct}
{\mcitedefaultendpunct}{\mcitedefaultseppunct}\relax
\EndOfBibitem
\bibitem[L{\"{a}}ngle \latin{et~al.}(2024)L{\"{a}}ngle, Mizohata, Mangler,
  Trentino, Mustonen, {\AA}hlgren, and Kotakoski]{Langle2024}
L{\"{a}}ngle,~M.; Mizohata,~K.; Mangler,~C.; Trentino,~A.; Mustonen,~K.;
  {\AA}hlgren,~E.~H.; Kotakoski,~J. {Two-dimensional few-atom noble gas
  clusters in a graphene sandwich}. \emph{Nature Materials} \textbf{2024},
  \emph{23}, 762--767\relax
\mciteBstWouldAddEndPuncttrue
\mciteSetBstMidEndSepPunct{\mcitedefaultmidpunct}
{\mcitedefaultendpunct}{\mcitedefaultseppunct}\relax
\EndOfBibitem
\bibitem[Vasu \latin{et~al.}(2016)Vasu, Prestat, Abraham, Dix, Kashtiban,
  Beheshtian, Sloan, Carbone, Neek-Amal, Haigh, Geim, and Nair]{Vasu2016}
Vasu,~K.~S.; Prestat,~E.; Abraham,~J.; Dix,~J.; Kashtiban,~R.~J.;
  Beheshtian,~J.; Sloan,~J.; Carbone,~P.; Neek-Amal,~M.; Haigh,~S.~J.;
  Geim,~A.~K.; Nair,~R.~R. {Van der Waals pressure and its effect on trapped
  interlayer molecules}. \emph{Nature Communications} \textbf{2016}, \emph{7},
  12168\relax
\mciteBstWouldAddEndPuncttrue
\mciteSetBstMidEndSepPunct{\mcitedefaultmidpunct}
{\mcitedefaultendpunct}{\mcitedefaultseppunct}\relax
\EndOfBibitem
\bibitem[K{\"u}hne \latin{et~al.}(2018)K{\"u}hne, B{\"o}rrnert, Fecher,
  Ghorbani-Asl, Biskupek, Samuelis, Krasheninnikov, Kaiser, and
  Smet]{kuhne2018reversible}
K{\"u}hne,~M.; B{\"o}rrnert,~F.; Fecher,~S.; Ghorbani-Asl,~M.; Biskupek,~J.;
  Samuelis,~D.; Krasheninnikov,~A.~V.; Kaiser,~U.; Smet,~J.~H. Reversible
  superdense ordering of lithium between two graphene sheets. \emph{Nature}
  \textbf{2018}, \emph{564}, 234--239\relax
\mciteBstWouldAddEndPuncttrue
\mciteSetBstMidEndSepPunct{\mcitedefaultmidpunct}
{\mcitedefaultendpunct}{\mcitedefaultseppunct}\relax
\EndOfBibitem
\bibitem[Lin \latin{et~al.}(2024)Lin, Matsumoto, Liu,
  Sol{\'{i}}s-Fern{\'{a}}ndez, Siao, Chiu, Ago, and Suenaga]{Lin2024nc}
Lin,~Y.-c.; Matsumoto,~R.; Liu,~Q.; Sol{\'{i}}s-Fern{\'{a}}ndez,~P.;
  Siao,~M.-d.; Chiu,~P.-w.; Ago,~H.; Suenaga,~K. {Alkali metal bilayer
  intercalation in graphene}. \emph{Nature Communications} \textbf{2024},
  \emph{15}, 425\relax
\mciteBstWouldAddEndPuncttrue
\mciteSetBstMidEndSepPunct{\mcitedefaultmidpunct}
{\mcitedefaultendpunct}{\mcitedefaultseppunct}\relax
\EndOfBibitem
\bibitem[Petrovi{\'{c}} \latin{et~al.}(2013)Petrovi{\'{c}}, {{\v{S}}rut
  Raki{\'{c}}}, Runte, Busse, Sadowski, Lazi{\'{c}}, Pletikosi{\'{c}}, Pan,
  Milun, Pervan, Atodiresei, Brako, {\v{S}}ok{\v{c}}evi{\'{c}}, Valla, Michely,
  and Kralj]{Petrovic2013}
Petrovi{\'{c}},~M.; {{\v{S}}rut Raki{\'{c}}},~I.; Runte,~S.; Busse,~C.;
  Sadowski,~J.~T.; Lazi{\'{c}},~P.; Pletikosi{\'{c}},~I.; Pan,~Z.-H.;
  Milun,~M.; Pervan,~P.; Atodiresei,~N.; Brako,~R.;
  {\v{S}}ok{\v{c}}evi{\'{c}},~D.; Valla,~T.; Michely,~T.; Kralj,~M. {The
  mechanism of caesium intercalation of graphene}. \emph{Nature Communications}
  \textbf{2013}, \emph{4}, 2772\relax
\mciteBstWouldAddEndPuncttrue
\mciteSetBstMidEndSepPunct{\mcitedefaultmidpunct}
{\mcitedefaultendpunct}{\mcitedefaultseppunct}\relax
\EndOfBibitem
\bibitem[{Al Balushi} \latin{et~al.}(2016){Al Balushi}, Wang, Ghosh,
  Vil{\'{a}}, Eichfeld, Caldwell, Qin, Lin, Desario, Stone, Subramanian, Paul,
  Wallace, Datta, Redwing, and Robinson]{AlBalushi2016}
{Al Balushi},~Z.~Y.; Wang,~K.; Ghosh,~R.~K.; Vil{\'{a}},~R.~A.;
  Eichfeld,~S.~M.; Caldwell,~J.~D.; Qin,~X.; Lin,~Y.~C.; Desario,~P.~A.;
  Stone,~G.; Subramanian,~S.; Paul,~D.~F.; Wallace,~R.~M.; Datta,~S.;
  Redwing,~J.~M.; Robinson,~J.~A. {Two-dimensional gallium nitride realized via
  graphene encapsulation}. \emph{Nature Materials} \textbf{2016}, \emph{15},
  1166--1171\relax
\mciteBstWouldAddEndPuncttrue
\mciteSetBstMidEndSepPunct{\mcitedefaultmidpunct}
{\mcitedefaultendpunct}{\mcitedefaultseppunct}\relax
\EndOfBibitem
\bibitem[P{\'{e}}cz \latin{et~al.}(2021)P{\'{e}}cz, Nicotra, Giannazzo,
  Yakimova, Koos, and Kakanakova‐Georgieva]{Pecz2021}
P{\'{e}}cz,~B.; Nicotra,~G.; Giannazzo,~F.; Yakimova,~R.; Koos,~A.;
  Kakanakova‐Georgieva,~A. {Indium Nitride at the 2D Limit}. \emph{Advanced
  Materials} \textbf{2021}, \emph{33}, 2006660\relax
\mciteBstWouldAddEndPuncttrue
\mciteSetBstMidEndSepPunct{\mcitedefaultmidpunct}
{\mcitedefaultendpunct}{\mcitedefaultseppunct}\relax
\EndOfBibitem
\bibitem[Forti \latin{et~al.}(2020)Forti, Link, St{\"{o}}hr, Niu, Zakharov,
  Coletti, and Starke]{Forti2020}
Forti,~S.; Link,~S.; St{\"{o}}hr,~A.; Niu,~Y.; Zakharov,~A.~A.; Coletti,~C.;
  Starke,~U. {Semiconductor to metal transition in two-dimensional gold and its
  van der Waals heterostack with graphene}. \emph{Nature Communications}
  \textbf{2020}, \emph{11}, 2236\relax
\mciteBstWouldAddEndPuncttrue
\mciteSetBstMidEndSepPunct{\mcitedefaultmidpunct}
{\mcitedefaultendpunct}{\mcitedefaultseppunct}\relax
\EndOfBibitem
\bibitem[Stark \latin{et~al.}(2019)Stark, Kuntz, Martens, and
  Warren]{Stark2019}
Stark,~M.~S.; Kuntz,~K.~L.; Martens,~S.~J.; Warren,~S.~C. {Intercalation of
  Layered Materials from Bulk to 2D}. \emph{Advanced Materials} \textbf{2019},
  \emph{31}, 1808213\relax
\mciteBstWouldAddEndPuncttrue
\mciteSetBstMidEndSepPunct{\mcitedefaultmidpunct}
{\mcitedefaultendpunct}{\mcitedefaultseppunct}\relax
\EndOfBibitem
\bibitem[Mustonen \latin{et~al.}(2022)Mustonen, Hofer, Kotrusz, Markevich,
  Hulman, Mangler, Susi, Pennycook, Hricovini, Richter, Meyer, Kotakoski, and
  Sk{\'{a}}kalov{\'{a}}]{Mustonen2022}
Mustonen,~K.; Hofer,~C.; Kotrusz,~P.; Markevich,~A.; Hulman,~M.; Mangler,~C.;
  Susi,~T.; Pennycook,~T.~J.; Hricovini,~K.; Richter,~C.; Meyer,~J.~C.;
  Kotakoski,~J.; Sk{\'{a}}kalov{\'{a}},~V. {Toward Exotic Layered Materials: 2D
  Cuprous Iodide}. \emph{Advanced Materials} \textbf{2022}, \emph{34},
  2106922\relax
\mciteBstWouldAddEndPuncttrue
\mciteSetBstMidEndSepPunct{\mcitedefaultmidpunct}
{\mcitedefaultendpunct}{\mcitedefaultseppunct}\relax
\EndOfBibitem
\bibitem[Yang \latin{et~al.}(2016)Yang, Li, Tang, Dong, Sun, Shen, and
  Wang]{yang2016sodium}
Yang,~S.; Li,~S.; Tang,~S.; Dong,~W.; Sun,~W.; Shen,~D.; Wang,~M. Sodium
  adsorption and intercalation in bilayer graphene from density functional
  theory calculations. \emph{Theoretical Chemistry Accounts} \textbf{2016},
  \emph{135}, 164\relax
\mciteBstWouldAddEndPuncttrue
\mciteSetBstMidEndSepPunct{\mcitedefaultmidpunct}
{\mcitedefaultendpunct}{\mcitedefaultseppunct}\relax
\EndOfBibitem
\bibitem[Lee and Persson(2012)Lee, and Persson]{lee2012li}
Lee,~E.; Persson,~K.~A. Li absorption and intercalation in single layer
  graphene and few layer graphene by first principles. \emph{Nano Letters}
  \textbf{2012}, \emph{12}, 4624--4628\relax
\mciteBstWouldAddEndPuncttrue
\mciteSetBstMidEndSepPunct{\mcitedefaultmidpunct}
{\mcitedefaultendpunct}{\mcitedefaultseppunct}\relax
\EndOfBibitem
\bibitem[Chepkasov \latin{et~al.}(2020)Chepkasov, Ghorbani-Asl, Popov, Smet,
  and Krasheninnikov]{chepkasov2020alkali}
Chepkasov,~I.~V.; Ghorbani-Asl,~M.; Popov,~Z.~I.; Smet,~J.~H.;
  Krasheninnikov,~A.~V. Alkali metals inside bi-layer graphene and MoS2:
  Insights from first-principles calculations. \emph{Nano Energy}
  \textbf{2020}, \emph{75}, 104927\relax
\mciteBstWouldAddEndPuncttrue
\mciteSetBstMidEndSepPunct{\mcitedefaultmidpunct}
{\mcitedefaultendpunct}{\mcitedefaultseppunct}\relax
\EndOfBibitem
\bibitem[Ji \latin{et~al.}(2019)Ji, Han, Hirata, Fujita, Shen, Ning, Liu,
  Kashani, Tian, Ito, Fujita, and Oyama]{ji2019lithium}
Ji,~K.; Han,~J.; Hirata,~A.; Fujita,~T.; Shen,~Y.; Ning,~S.; Liu,~P.;
  Kashani,~H.; Tian,~Y.; Ito,~Y.; Fujita,~J.-i.; Oyama,~Y. Lithium
  intercalation into bilayer graphene. \emph{Nature Communications}
  \textbf{2019}, \emph{10}, 275\relax
\mciteBstWouldAddEndPuncttrue
\mciteSetBstMidEndSepPunct{\mcitedefaultmidpunct}
{\mcitedefaultendpunct}{\mcitedefaultseppunct}\relax
\EndOfBibitem
\bibitem[K{\"u}hne \latin{et~al.}(2017)K{\"u}hne, Paolucci, Popovic, Ostrovsky,
  Maier, and Smet]{kuhne2017}
K{\"u}hne,~M.; Paolucci,~F.; Popovic,~J.; Ostrovsky,~P.~M.; Maier,~J.;
  Smet,~J.~H. Ultrafast lithium diffusion in bilayer graphene. \emph{Nature
  Nanotechnology} \textbf{2017}, \emph{12}, 895--900\relax
\mciteBstWouldAddEndPuncttrue
\mciteSetBstMidEndSepPunct{\mcitedefaultmidpunct}
{\mcitedefaultendpunct}{\mcitedefaultseppunct}\relax
\EndOfBibitem
\bibitem[Li \latin{et~al.}(2024)Li, Börrnert, Ghorbani-Asl, Biskupek, Zhang,
  Zhang, Bresser, Krasheninnikov, and Kaiser]{YueliangLi24}
Li,~Y.; Börrnert,~F.; Ghorbani-Asl,~M.; Biskupek,~J.; Zhang,~X.; Zhang,~Y.;
  Bresser,~D.; Krasheninnikov,~A.~V.; Kaiser,~U. In Situ TEM Investigation of
  the Lithiation and Delithiation Process Between Graphene Sheets in the
  Presence of Atomic Defects. \emph{Advanced Functional Materials}
  \textbf{2024}, \emph{n/a}, 2406034\relax
\mciteBstWouldAddEndPuncttrue
\mciteSetBstMidEndSepPunct{\mcitedefaultmidpunct}
{\mcitedefaultendpunct}{\mcitedefaultseppunct}\relax
\EndOfBibitem
\bibitem[Zhang \latin{et~al.}(2023)Zhang, Ghorbani-Asl, Zhang, and
  Krasheninnikov]{Zhang2023Li}
Zhang,~X.; Ghorbani-Asl,~M.; Zhang,~Y.; Krasheninnikov,~A.~V. {Quasi-2D FCC
  lithium crystals inside defective bi-layer graphene: insights from
  first-principles calculations}. \emph{Materials Today Energy} \textbf{2023},
  \emph{34}, 101293\relax
\mciteBstWouldAddEndPuncttrue
\mciteSetBstMidEndSepPunct{\mcitedefaultmidpunct}
{\mcitedefaultendpunct}{\mcitedefaultseppunct}\relax
\EndOfBibitem
\bibitem[Liu \latin{et~al.}(2014)Liu, Kutana, Liu, and Yakobson]{Liu2014Yak}
Liu,~M.; Kutana,~A.; Liu,~Y.; Yakobson,~B.~I. {First-Principles Studies of Li
  Nucleation on Graphene}. \emph{The Journal of Physical Chemistry Letters}
  \textbf{2014}, \emph{5}, 1225--1229\relax
\mciteBstWouldAddEndPuncttrue
\mciteSetBstMidEndSepPunct{\mcitedefaultmidpunct}
{\mcitedefaultendpunct}{\mcitedefaultseppunct}\relax
\EndOfBibitem
\bibitem[Chepkasov \latin{et~al.}(2022)Chepkasov, Smet, and
  Krasheninnikov]{chepkasov2022single}
Chepkasov,~I.~V.; Smet,~J.~H.; Krasheninnikov,~A.~V. Single-and Multilayers of
  Alkali Metal Atoms inside Graphene/MoS2 Heterostructures: A Systematic
  First-Principles Study. \emph{The Journal of Physical Chemistry C}
  \textbf{2022}, \emph{126}, 15558--15564\relax
\mciteBstWouldAddEndPuncttrue
\mciteSetBstMidEndSepPunct{\mcitedefaultmidpunct}
{\mcitedefaultendpunct}{\mcitedefaultseppunct}\relax
\EndOfBibitem
\bibitem[Moriwake \latin{et~al.}(2017)Moriwake, Kuwabara, Fisher, and
  Ikuhara]{moriwake2017sodium}
Moriwake,~H.; Kuwabara,~A.; Fisher,~C.~A.; Ikuhara,~Y. Why is
  sodium-intercalated graphite unstable? \emph{RSC Advances} \textbf{2017},
  \emph{7}, 36550--36554\relax
\mciteBstWouldAddEndPuncttrue
\mciteSetBstMidEndSepPunct{\mcitedefaultmidpunct}
{\mcitedefaultendpunct}{\mcitedefaultseppunct}\relax
\EndOfBibitem
\bibitem[Liu \latin{et~al.}(2016)Liu, Merinov, and Goddard]{liu2016origin}
Liu,~Y.; Merinov,~B.~V.; Goddard,~W.~A. Origin of low sodium capacity in
  graphite and generally weak substrate binding of Na and Mg among alkali and
  alkaline earth metals. \emph{Proceedings of the National Academy of Sciences}
  \textbf{2016}, \emph{113}, 3735--3739\relax
\mciteBstWouldAddEndPuncttrue
\mciteSetBstMidEndSepPunct{\mcitedefaultmidpunct}
{\mcitedefaultendpunct}{\mcitedefaultseppunct}\relax
\EndOfBibitem
\bibitem[Su \latin{et~al.}(2019)Su, Prestat, Hu, Puthiyapura, Neek-Amal, Xiao,
  Huang, Kravets, Haigh, Hardacre, Peeters, and Nair]{Su2019}
Su,~Y.; Prestat,~E.; Hu,~C.; Puthiyapura,~V.~K.; Neek-Amal,~M.; Xiao,~H.;
  Huang,~K.; Kravets,~V.~G.; Haigh,~S.~J.; Hardacre,~C.; Peeters,~F.~M.;
  Nair,~R.~R. {Self-Limiting Growth of Two-Dimensional Palladium between
  Graphene Oxide Layers}. \emph{Nano Letters} \textbf{2019}, \emph{19},
  4678--4683\relax
\mciteBstWouldAddEndPuncttrue
\mciteSetBstMidEndSepPunct{\mcitedefaultmidpunct}
{\mcitedefaultendpunct}{\mcitedefaultseppunct}\relax
\EndOfBibitem
\bibitem[Araki \latin{et~al.}(2022)Araki, Sol{\'{i}}s-Fern{\'{a}}ndez, Lin,
  Motoyama, Kawahara, Maruyama, Gao, Matsumoto, Suenaga, Okada, and
  Ago]{Araki2022acs}
Araki,~Y.; Sol{\'{i}}s-Fern{\'{a}}ndez,~P.; Lin,~Y.~C.; Motoyama,~A.;
  Kawahara,~K.; Maruyama,~M.; Gao,~Y.; Matsumoto,~R.; Suenaga,~K.; Okada,~S.;
  Ago,~H. {Twist Angle-Dependent Molecular Intercalation and Sheet Resistance
  in Bilayer Graphene}. \emph{ACS Nano} \textbf{2022}, \emph{16},
  14075--14085\relax
\mciteBstWouldAddEndPuncttrue
\mciteSetBstMidEndSepPunct{\mcitedefaultmidpunct}
{\mcitedefaultendpunct}{\mcitedefaultseppunct}\relax
\EndOfBibitem
\end{mcitethebibliography}

 \end{document}